# Molecular magnetism and crystal field effects in the Kondo system Ce₃Pd₂₀(Si,Ge)₆ with two Ce sublattices

V.N. Nikiforov[a], Yu.A. Koksharov[a,b], A.V. Gribanov[a], M. Baran[c], V.Yu. Irkhin[d],*

[a]*M.V. Lomonosov Moscow State University, Moscow, 119992, Russia*
[b]*Kurnakov Institute of General and Inorganic Chemistry, Leninskii pr. 31, Moscow, 119991 Russia*
[c]*Institute of Physics, Polish Academy of Science, Warsaw, Poland*
[d]*Institute of Metal Physics, Ekaterinburg, 620990, Russia*

## Abstract

The unusual electronic and magnetic properties of the systems Ce₃Pd₂₀T₆ (T = Ge, Si) with two non-equivalent Ce positions are discussed. The logarithmic growth of resistance for both systems confirms the presence of the Kondo effect in the two respective temperature ranges. The two-scale behavior is explained by consecutive splitting of Ce ion levels in the crystal field. The effects of the frustration caused by the coexistence of the different positions of cerium are treated, which may also significantly enhance the observed values of specific heat. A model of "molecular magnetism" with Ce2 cubes is developed..

## 1. Introduction

The Kondo systems Ce₃Pd₂₀T₆ (with T = Ge, Si) have unique electronic properties [1-4]. The interesting feature of this cubic system is two nested cubes composed of cerium ions, Ce1 and Ce2, in two non-equivalent positions [3,4]. These systems have a high electron heat capacity: estimations above T = 6 K give the value of γ = 0.21 and 0.3 J / mol K² for T = Si and Ge. Ce₃Pd₂₀Si₆ yield at ultralow temperatures anomalous heat capacity which is almost constant at very low temperatures 0.3 K÷2K, and formally calculated γ grows up to about $10^4$ mJ/mol K². The structural features (high symmetry with cluster motifs) lead to additional peculiarities of electronic and magnetic properties.

First magnetic and transport properties of the 3-20-6 system (in particular, manifestations of the Kondo effect) were reported in 1994 [1,2]. After discussion of this report, the paper [3] was published. In detail, magnetic and electrical properties have been investigated in [3,4], where a large role of magnetic frustrations caused by features of the lattice (cluster "superstructure") was revealed.

Studies of the magnetic structure were performed in [5]. A model of quadrupolar ordering below 1 K proposed for Ce₃Pd₂₀(Si,Ge)₆ [6]. However, this model apparently does not describe the whole variety of anomalies in the systems discussed. At the same time, the "molecular magnetism" approach in the framework of the crystal field (CEF) theory [3-4] allows us to describe a broader set of properties in the range from 2K to room temperatures. This analysis is the focus of the present work.

## 2. Experimental and structure

For the first time, Ce₃Pd₂₀Si₆ and Ce₃Pd₂₀Ge₆ single phase samples were synthesized at the Chemistry Department of Moscow State University. These samples have a high quality, and the properties were found to be consistent with the literature data. X-ray analysis carried out by group (Yu.D. Seropegin, Moscow) revealed that both the compounds have a cubic symmetry (space group *Fm-3m*) and correspond to the Mg₃Ni₂₀B₆ structural type which is an ordered derivative from the binary Cr₂₃C₆ type [7]. The most intriguing fact is the existence of two non-equivalent positions of the RE elements (Fig.1).

*E-mail address*: valentin.irkhin@imp.uran.ru
.

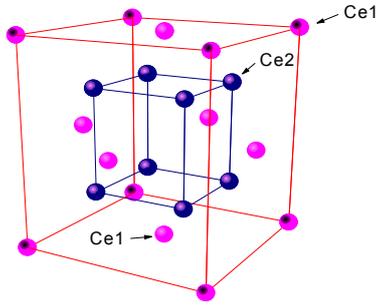

Fig.1. Crystal structure of Ce$_3$Pd$_{20}$T$_6$

Nowadays, there are a number of compounds belonging to this superstructure, e.g., silicides R$_3$Pd$_{20}$Si$_6$ (R= La, Ce, Sm, Yb) and germanide Ce$_3$Pd$_{20}$Ge$_6$. Measurements were made for many 4f-compounds (except for La$_3$Pd$_{20}$Si$_6$), which allowed to explore the replacement of the rare-earth and the third element.

Investigations of the temperature dependence of the resistivity, magnetic susceptibility, specific heat, thermal conductivity, Seebeck coefficient for a large group of the synthesized samples were performed. The heat capacity and transport properties were measured by standard dc four probe techniques on Quantum Design Physical Property Measurement System. The static magnetic moment was measured by using the vibrating sample magnetometer PARC-M155 and SQUID magnetometer Quantum Design.

### 3. Results on electronic and magnetic properties

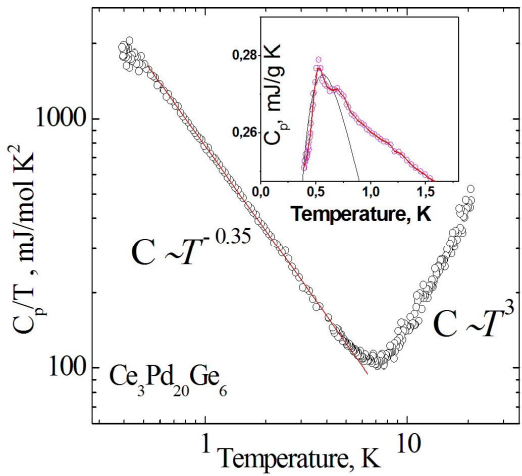

Fig.2. Temperature dependencies of specific heat al low temperatures. The inset shows the ultralow temperature region, the line being a Schottky-like fit.

The specific heat measurements performed on Ce$_3$Pd$_{20}$Ge$_6$ down to 0.3 K yield a giant value of the Sommerfeld coefficient $\gamma$ = 3 J/mol K$^2$. However, calorimetric experiments at 4–25K show moderate $\gamma$ about 300 mJ/mol K$^2$. One can see from inset in Fig. that $\gamma$ starts to decrease at low $T$. The dependence $T^{-0.35}$ seems to indicate a spin-glass-like behavior.

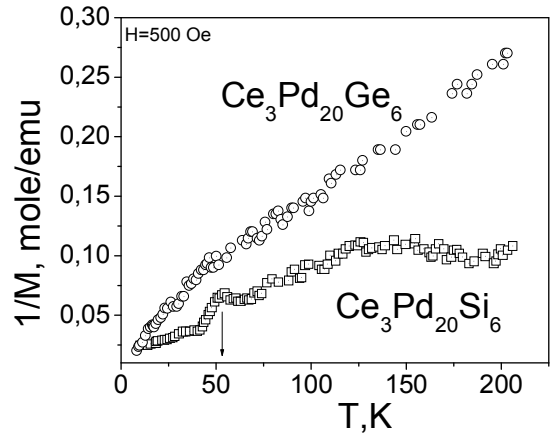

Fig.3. Temperature dependencies of inverse magnetization in a weak field H = 500 Oe

Ultralow-temperature anomalies in $C(T)$ can be due to antiferromagnetic ordering and quadrupole transition [6]. Besides that, a local symmetry lowering of the $\Gamma_8$ ground state and CEF level splitting may play a role.

Temperature dependences of magnetization in a small field and of resistivity are presented in Figs.3, 4.

A fundamental difference between the behavior of $M(T)$ in a field of 500 Oe for Ce$_3$Pd$_{20}$Si$_6$ and Ce$_3$Pd$_{20}$Ge$_6$ is the existence of anomalies of rather complicated form for Ce$_3$Pd$_{20}$Si$_6$ in the range 40-50 K, indicating a change in the magnetic structure of this compound. This anomaly is not typical for both ferro- and antiferromagnetic transition in all the spin structure of the compound, but seems to be due to the existence of a two-sublattice spin structure It is clearly seen that this anomaly is imposed on the normal behavior of $M(T)$ in the case of the Curie-Weiss paramagnetism. It should be noted that in this temperature range the transition from the linear to nonlinear dependence M($H$) in low fields.

For all the samples, picking out the magnetic part $\rho_M(T)$ of resistivity was carried out by subtracting the resistance of the phonon contribution. The logarithmic growth of resistivity for both the systems confirms the presence of the Kondo effect in the two respective temperature ranges. Namely, the temperature corresponding to the resistance minimum is $T_{min}$ = 71K and 12.5 K, and $T_{max}$ = 332K and 151K for silicon and germanium systems respectively.

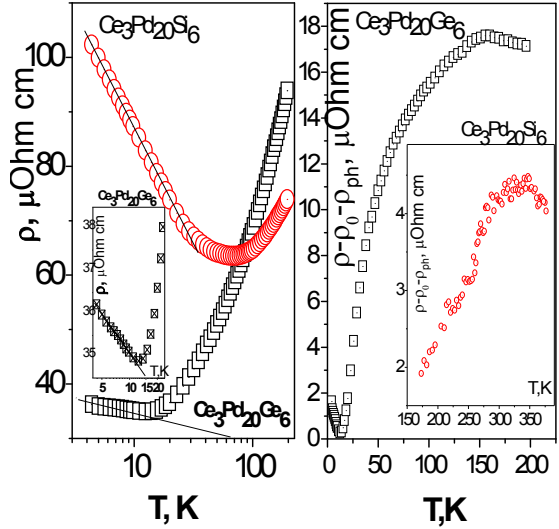

Fig.4.Temperature dependencies of resistivity al low temperature (left) and at more high temperatures (right, the phonon contribution is subtracted according to the Blokh-Gruneisen formula, the Debye temperature being obtained from $T^3$ specific heat contribution)

Magnetic resistance behavior $\rho_M(T)$ for both samples (Fig.4) is different. For $Ce_3Pd_{20}Si_6$ at high temperatures $\rho_M$ varies slightly, but upon cooling grows logarithmically, while for $Ce_3Pd_{20}Ge_6$ $\rho_M$ decreases with decreasing temperature, and below 10 K logarithmic increase starts too. When approaching the temperature $T$ = 2 K, $\rho_M(T)$ for $Ce_3Pd_{20}Ge_6$ reaches a maximum, possibly corresponding to the transition into "correlated scattering" regime. The logarithmic singularity for $Ce_3Pd_{20}Si_6$ is much larger than for $Ce_3Pd_{20}Ge_6$, which indicates a higher Kondo temperature.

Another important result is that for both samples $\rho_M(T)$ shows no anomalies at the magnetic transition temperature. Unlike Sm compounds, where there is a sharp drop in resistance at $T_N$, in both the samples a pronounced logarithmic Kondo growth resistance takes place, which has usually a strong sensitivity with respect to magnetic ordering. This also evidences in favor of the assumption that the Kondo effect occurs only for some subsystem of Ce sites, while the other subsystem is exposed to magnetic ordering.

## 4. Crystal field effects

The ground state $^2F_{5/2}$ of an isolated $Ce^{3+}$ ion is six-fold degenerate. The ground state ($^2F_{5/2}$) and the next excited ($^2F_{7/2}$) are very distant in energy (about 3100 K). For metallic cerium compounds the CEF interaction results in a total splitting of the ground-state J-multiplet of typically $10^1$-$10^2$ K. Since the CEF splitting of the ground-state J-multiplet is small as compared to the distance to the first excited J-multiplet, the CEF is treated usually as the perturbation within the ground-state J-multiplet only.

In the case of cubic crystal field $^2F_{5/2}$ ground state of the $Ce^{3+}$ ion is split into doublet ($\Gamma_7$) and quartet ($\Gamma_8$). These $\Gamma_7$ and $\Gamma_8$ states have markedly different charge clouds. The point-charge model of the CEF interaction predicts the simple rule: if the crystal field arises from an octahedron (a cube) of six (eight) surrounding negative charges, then the low-lying level is $\Gamma_7$ ($\Gamma_8$). The ordering of $\Gamma_7$ and $\Gamma_8$ levels in cerium-based Kondo compounds can be the subject of controversy. In most intermetallic cerium compounds $Ce^{3+}$ ion has the CEF ground state $\Gamma_7$. However, e.g., in CeAg and $CeB_6$, the ground state is the $\Gamma_8$ quadruplet. The temperature dependencies $\chi(T)$ for $\Gamma_7$ and $\Gamma_8$ ground states of a $Ce^{3+}$ ion can be rather easily discriminated (Fig.5).

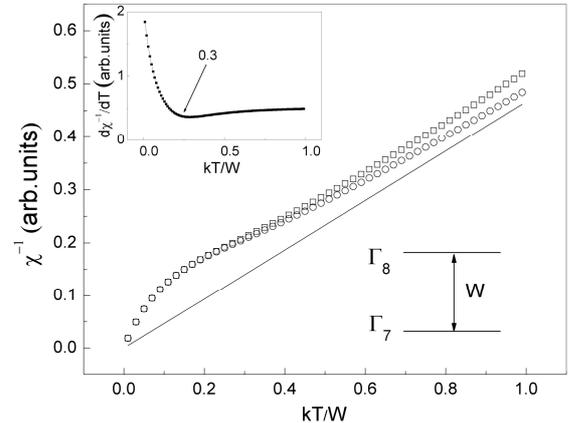

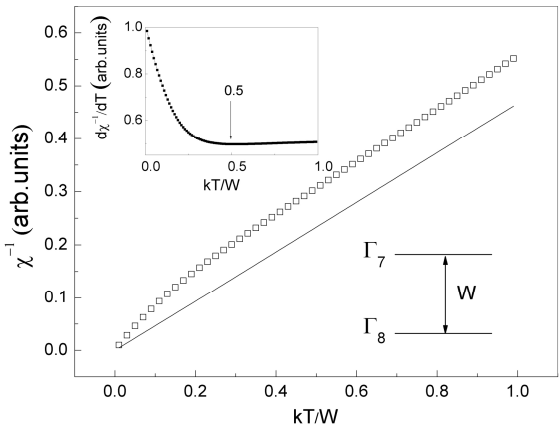

Fig.5. Inverse magnetic susceptibility for $Ce^{3+}$ ($^2F_{5/2}$) ions with $\Gamma_7$ and $\Gamma_8$ ground states. Squares are fit according to [T8], circles according to [9]. Solid line is the Curie-Weiss law for free $Ce^{3+}$ ($^2F_{5/2}$) ion. Inserts present the corresponding $d\chi^{-1}/dT$ curves, he minimum being shown.

The cerium ions in $Ce_3Pd_{20}Ge_6$ and $Ce_3Pd_{20}Si_6$ have two different positions (Ce1 and Ce2) [7]. Coordination polyhedra of the cerium atoms in Ce1 (4a) position are cubo-octahedrons and in Ce2 (8c) position are normal polyhedra with 16 apexes. In the early work [4] the temperature dependence of the magnetic susceptibility in $Ce_3Pd_{20}Ge_6$ indicated to $\Gamma_7$ as the ground state of both $Ce^{3+}$ ions with $W = 250$-$450$ K. However, the low-temperature specific heat measurements [6] showed possible quadrupolar and antiferromagnetic ordering, similar $CeB_6$. Later the INS experiments revealed excitations at 60 K and 46 K (corresponding to Ce1 and Ce2 positions) and demonstrated that in $Ce_3Pd_{20}Ge_6$ all cerium ions have $\Gamma_8$ as the ground state [10]. In inelastic neutron scattering on $Ce_3Pd_{20}Si_6$, excitations at 45 K (related to Ce1) and 3.6 K (related to Ce2) were resolved [1]. It is interesting that according to [11] the ground states of the two cerium sites in $Ce_3Pd_{20}Si_6$ are non-equivalent: Ce1 and Ce2 have a $\Gamma_7$ and $\Gamma_8$ ground state, respectively. This is in contrast to the situation found in the analogous compound $Ce_3Pd_{20}Ge_6$, where the $\Gamma_8$ ground state is found for both Ce sites. The results of [11] demonstrate also an interaction between the 8c and 4a sites previously believed to be absent..

There are some problems with interpretation of the CEF schemes in $Ce_3Pd_{20}Ge_6$ and $Ce_3Pd_{20}Si_6$. In particular, the results of [11] were critically analyzed in [12]. Probably, in INS experiments on $Ce_3Pd_{20}Si_6$ only one excitation (at 45 K) is unambiguously resolved. Since the magnetic entropy exceeds $3R\ln2$ per formula unit at 1.5 K, if the excitation energy is 45 K for both Ce sites, this would imply that at least one of the Ce sites assumes a $\Gamma_8$ ground state.

Further on, the temperature dependences of magnetic susceptibility should have marked CEF features at 15-30 K. Most detailed experimental curves of $\chi(T)$ [6] do not reveal those. Finally, the CEF scheme should correlate with the crystal structure of compounds. For example, the Ce2 coordination polyhedra, including 16 Pd atoms, are nearly identical in germanium and siliceous systems, while the Ce1 coordination polyhedra changes significantly [7]. From point-charge CEF considerations this implies that the INS excitation at $\approx 45$ K, observed both in $Ce_3Pd_{20}Ge_6$ and $Ce_3Pd_{20}Si_6$, could be related to Ce2 position, as it is confirmed for $Ce_3Pd_{20}Ge_6$ in [10] and obscure for $Ce_3Pd_{20}Si_6$.

## 5. Molecular magnetism

The Ce2 ions in the crystal structure of $Ce_3Pd_{20}Ge_6$ and $Ce_3Pd_{20}Si_6$ compose "small" cube included into "big" cube of Ce1 [4]. Some experimental evidences indicate a possibility of relatively weak interaction between Ce1 and Ce2 ions: the existence of two Kondo temperatures, and different behavior of Ce1 and Ce2 at low-temperature quadrupole and antiferromagnetic transitions [5]. Probably, to some extent one can consider Ce1 and Ce2 as separate magnetic subsystems.

Presence of two Ce subsystems in the cubic structure 3-20-6 enables us to suppose that one of these subsystems contains moments which undergo magnetic ordering, while the second is responsible for the ordinary paramagnetic properties. Apparently, only the moments of nested Ce2 subcube can be ordered, since the atoms Pd are located just inside this, while in outer space (i.e., between the cubes Ce2) are all the rest atoms (Ce1, Pd and Si). Thus, these "cubes" are sufficiently isolated. In turn, Ce1 cubes contain nested cubes Ce2, which makes more than doubtful a cube model with Ce1 as ordered moments excluding Ce2. So, we assume that Ce2 forms some magnetic molecule, its magnetic moment increasing with decreasing temperature.

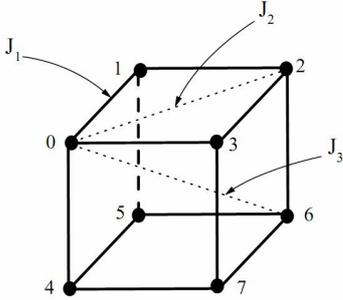

Fig.6. A sketch of a cubic cluster with eight spins. $J_i$ ($i = 1\div3$) denote the exchange constants.

Hence, we can consider the "small" cube of Ce2 as an exchange-coupled cluster. Such cubic cluster can have unusual magnetic properties [13].

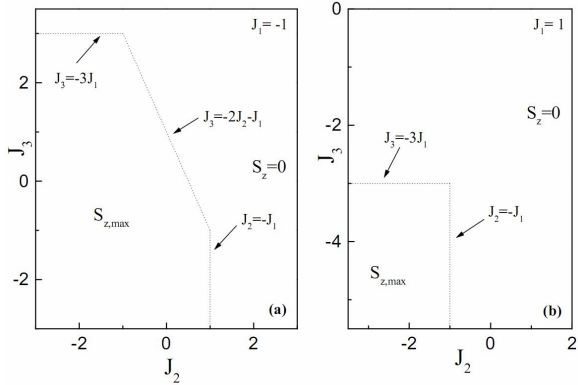

Fig.7. The universal "phase diagrams" for the Ising cubic cluster. Dotted lines show phase boundaries between the ground states with the spin projections $S_{z,max}$ (±4 for $s=1/2$, ±8 for $s=1$, etc) and $S_z=0$.

In magnetic clusters, each of spin has a few neighbors, resulting in a frustration of exchange interactions. Because of complexity of exchange pathways, it is difficult to predict *a priori* both the sign and magnitude of exchange constants. A cluster becomes magnetic, if it has a non-zero-spin ground state. It is interesting to study effects of competing interactions on the ground state of a relatively small and highly symmetrical spin cluster, for which a wide range of exchange parameters can be relatively easily examined. Here we consider magnetic properties (ground states) of cubic clusters of eight spins s, coupled by Ising or Heisenberg interactions. A sketch of such cluster with possible exchange pathways is shown in Fig.6.

The "phase diagrams" in Fig.7 indicates that complex exchange interactions inside Ce2 cube in $Ce_3Pd_{20}Ge_6$ and $Ce_3Pd_{20}Si_6$ could result in zero total magnetic moment

($S=0$). Moreover, for antiferromagnetic exchange interactions $J_1>0$ the non-magnetic ground state seems to be mostly probable. It is interesting that in $Ce_3Pd_{20}Ge_6$ a possible ordered cerium moment at the 8c site is found to be reduced to an undetectable small value down to the lowest measured temperature T= 0.05 K [5]. Similarly, in $Ce_3Pd_{20}Si_6$ cerium ions are magnetically ordered only in (4a) positions [11]. Probably, a molecular-type magnetism of Ce2 cube promotes non-magnetic ground state of the (8c) subsystem at low temperatures. It is important that the Ising "phase diagram" (Fig.3) can be used as a rough approximation for the Heisenberg model

Frustration of exchange interaction can result in suppression of magnetic ordering and tendency to formation of a spin-liquid state with unusual properties [14]. The effects of the frustrations, together with the Kondo anomalies, may also increase significantly enhance the observed values of the specific heat in a metallic antiferromagnet [14,15].

## 6. Discussion and conclusions

We have demonstrated that the system $Ce_3Pd_{20}(Ge,Si)_6$ is one of few cerium compounds with different crystallographic positions and a clear separation of the two salient trends in physics of 4f-intermetallics: the Kondo effect and magnetic ordering.

Magnetic measurements support the assumption that cerium atoms in positions Ce1 are responsible for the manifestation of the Kondo effect in the both compounds. In $Ce_3Pd_{20}Ge_6$ the nearest neighbors of Ce1 are Pd atoms, while for $Ce_3Pd_{20}Si_6$ they are Si atoms which are located much closer than the sum of the radii of metallic Pd and Ce atoms [7]. However, it is possible that a similar Kondo effect takes place for the Ce2 substructure 3-20-6.

The two-scale behavior of resistivity can be explained by consecutive splitting of Ce ion levels in the crystal field. The hierarchy of these scales (the Kondo temperatures) corresponds to the hierarchy of CEF parameters, which can be treated in spirit of the consideration [16].

For germanium system, the role of the Kondo effect is much smaller than in $Ce_3Pd_{20}Si_6$, which correspondingly reduces the scale of the observed anomalies. Replacing of Ce by U or Sm (which will be discussed elsewhere) leads to a dramatic disappearance of anomalies under consideration.

The paper [17] discusses the physics of a Kondo lattice model with two local-moment sublattices, coupled with different Kondo couplings to conduction electrons with application to $Ce_3Pd_{20}(Si,Ge)_6$. The phase diagram will be strongly modified from that of the standard Kondo lattice if the characteristic screening temperatures of the distinct moments are well separated.

Authors of [18] identify a cubic heavy-fermion material including $Ce_3Pd_{20}Si_6$ as exhibiting a field-induced quantum phase transition. The transition between two different ordered phases is accompanied by an abrupt change of Fermi surface.

CEF effects in the Kondo-lattice compounds $Ce_3Pd_{20}Ge_6$ and $Ce_3Pd_{20}Si_6$ require further studying since some serious questions remain unanswered. In our opinion, the main problem is matching of results of inelastic neutron scattering experiments and temperature dependences of magnetic susceptibility. Another problem is ambiguous CEF scheme for Ce1 and Ce2 ions in $Ce_3Pd_{20}Si_6$. The mechanisms and intensity of interactions between two cerium subsystem (Ce1 and Ce2) are also obscure. Coexistence and competing of various physics phenomena - Kondo scattering and screening, magnetic and possible orbital (quadrupolar) ordering, phonon anomalies (rattling, etc.) and quantum phase transitions – makes $Ce_3Pd_{20}Ge_6$ and $Ce_3Pd_{20}Si_6$ very interesting and nontrivial objects for researchers.

**Acknowledgements**

This work is supported in part by the grant No 14-02-92019 from Russian Foundation for Basic Research, by Programs of RAS Physical Division, project No. 12-T-2-1001 (Ural Branch) and of RAS Presidium, project No. 12-P-2-1041.

1. V.N. Nikiforov et al, Int. Conf. on Solid Compound of Transition Elements, Wroclav, 1994, pp.79-90; Int. Conf. Magnetism, Warsaw, ICM 1994, Poland, 1994, p.557.
2. N. Takeda, J. Kitagawa and M. Ishikawa, :J. Phys. Soc. Jpn. 64 (1995) 387.
3. Yu P. Gajdukov, Yu.A. Koksharov, J. Mirković, Yu.V. Kochetkov, V.N. Nikiforov, JETP Letters, 61 (1995) 391.
4. V.N. Nikiforov, Yu.A. Koksharov, Yu.V. Kochetkov, J. Mirković, J. Magn. Magn. Mat., 163 (1996) 184.
5. A. Donni, T. Herrmannsdörfer, P. Fischer, L .Keller, F .Fauthr, K.A. McEven, T. Goto, T. Komatsubara, J. Phys:: Cond. Mat. 12 (2000) 9441
6. J. Kitagawa, N. Takeda and M. Ishikawa, Phys.Rev.B 53 (1996) 5101
7. A. V. Gribanov, Yu. D. Seropegin and O. I. Bodak, J. Alloys Compd. 204 (1994) L9
8. C. Terzioglu, D.A. Browne, R.G. Goodrich, A. Hassan, Z. Fisk, Phys.Rev.B 63 (2001) 235110
9. T. Murao, T. Matsubara, Progr. Theor. Phys. 18 (1957) 215
10. L. Keller, A. Dönni, M. Zolliker, and T. Komatsubara, Physica B 259-261 (1999) 336
11. P.P. Deen, A.M. Strydom, S. Paschen, D.T. Adroja, W. Kockelmann, S.Rols, Phys.Rev.B 81 (2010) 064427
12. S. Pashen, J. Larrea,. arXiv:1403.1386v1
13. Yu.A. Koksharov, arXiv:cond-mat/0112475v1
14. S.V. Vonsovsky, V.Yu. Irkhin, M.I. Katsnelson, Physica B171 (1991) 135
15. V. Yu. Irkhin, M. I. Katsnelson, Phys.Rev. B 62 (2000) 5647
16. S. Kashiba, S. Maekawa, S. Takahashi, and M. Tachiki, J. Phys. Soc. Jpn 55 (1986) 1341
17. A. Benlagra, L. Fritz, M. Vojta, Phys. Rev. B 84, 075126 (2011)
18. J. Custers, K-A. Lorenzer, M.Müller, A. Prokofiev, A. Sidorenko, H.Winkler, A.M. Strydom, Y. Shimura, T. Sakakibara, R. Yu, Q. Si, S.Paschen, Nature Materials 11 (2012) 189